\documentclass[prm,twocolumn, amsmath,amssymb,
reprint,%
author-year,%
author-numerical,%
dvipdfmx,
superscriptaddress
]{revtex4-1}
\usepackage{graphicx}% Include figure files
\usepackage{bm}% bold math
\usepackage{color}
\begin{document}

%\preprint{APS/123-QED}
 
\title{Composition-dependent ultrafast luminescence in Cu-Ni alloys: Combined experimental and ab initio study}

\author{Tohru Suemoto}
\email{suemoto@issp.u-tokyo.ac.jp}
\affiliation{Department of Engineering Science, The University of Electro-Communications, Chofu, Tokyo 182-8585, Japan}

\author{Haruki Morino}
\affiliation{Department of Engineering Science, The University of Electro-Communications, Chofu, Tokyo 182-8585, Japan}

\author{Shota Ono}
\email{shotaono@muroran-it.ac.jp}
\affiliation{Department of Sciences and Informatics, Muroran Institute of Technology, Muroran 050-8585, Japan}

\author{Tsuyoshi Okuno}
\affiliation{Department of Engineering Science, The University of Electro-Communications, Chofu, Tokyo 182-8585, Japan}

\author{Takeshi Suzuki}
\author{Kozo Okazaki}
\author{Shuntaro Tani}
\author{Yohei Kobayashi}
\affiliation{Institute for Solid State Physics, The University of Tokyo, Kashiwanoha 5-1-5, Kashiwa, Chiba, Japan}

\begin{abstract}
Properties of Cu-Ni solid solutions have long been studied in physical and materials sciences. Yet, their many-body properties have not been well understood. Here, we investigate ultrafast luminescence in near infrared region for Cu$_{1-x}$Ni$_x$ alloys. The luminescence intensity was the highest in Cu and decreased dramatically by adding Ni, approaching close to the value for pure Ni at $x=0.45$. This composition dependence was well reproduced by calculations assuming two body scattering of the energetic electrons. The luminescent decay rate was not straightforward, i.e., it decreased first by adding Ni up to $x=0.17$ and then started to increase approaching twice the initial value at $x=0.45$. This behavior was in good agreement with ab initio calculations of electron-phonon (e-ph) coupling strength. This work provides a new perspective on the electron relaxation dynamics in solid solutions systems. 
\end{abstract}
%\pacs{}

\maketitle

%{\color{blue} One of the strategies for material design is element substitution for a base material. It is important to study which kind of atoms is added and what amount of atoms is added. }

%%%%%%%%%%%%%%%%%%%%%%%%%%%%%%%%%%%%%%%%%%%%%%%%%%%
{\it Introduction---}Metal alloys are of significant importance both in application and solid state physics, because the crystal structure and phase stability \cite{vegard,calphad}, electrical and thermal conductivity \cite{bhatia,ying,wei}, magnetization \cite{james,iwasaki}, and optical properties \cite{gong} can be greatly modified from the constituent metals by alloying. In addition to these macroscopic properties, it is of fundamental interest to understand the excited state dynamics and the mutual interactions between electrons and phonons. Many body properties in an ordered compound have been investigated in detail \cite{giri}, but those in solid solutions are not well understood.

%Recently, the dynamics of energetic electrons in alloy attract considerable attention for optimization of laser machining condition \cite{someone}. 

Cu-Ni alloy is a traditional solid solution system, utilized for coins with silver-like color for a long time. The nearly temperature independent resistivity and low thermoelectric power in Cu$_{0.55}$Ni$_{0.45}$ alloy (so-called constantan) are utilized for thermocouples (known as copper-constantan) and standard register. Relatively low thermal conductivity in cupronickel (Cu$_{0.7}$Ni$_{0.3}$) is useful for cryogenic instrument. The advantage of this system is that Cu and Ni have the same crystal structure (FCC) with lattice constant close each other, ensuring homogeneous mixture at arbitrary composition. 
The lattice constant of Cu$_{1-x}$Ni$_x$ alloys linearly decreases with the Ni concentration $x$, satisfying the well-known Vegard's law \cite{vegard}. However, their magnetic and electronic properties exhibit more complex behaviors with increasing $x$. The Cu-Ni alloys show no magnetization below $x \simeq x_c = 0.44$, above which the ferromagnetic order starts to increase \cite{ahern,ikeda}. The resistivity of Cu-Ni alloy is far larger than the end materials \cite{ikeda,ahmad,paras}. This nonmonotonic composition dependence of the residual resistivity is ascribed to impurity scattering \cite{ikeda} and the magnetic resistivity due to the electron scattering from spin antiparallel to parallel states \cite{kasuya}. Although the electron-electron (e-e) and electron-phonon (e-ph) interaction may contribute to the excited state dynamics, their composition-dependence of Cu-Ni alloys is not well understood.

%. For example, the alloy with $x=0.2$ is 3 times larger than pure Ni and 13 times larger than pure Cu at room temperature \cite{paras}

The ultrafast relaxation dynamics following a pulsed laser excitation enables us to study the e-e and e-ph properties of solids. Recently, we have systematically studied ultrafast luminescence of various elemental metals \cite{suemoto2019,suemoto2020,suemoto2021,suemoto2023,suemoto2025}. The time evolution of the luminescence intensity can be decomposed into a quasi-instantaneous and an exponential decay component. We have shown that (i) the intensity of the quasi-instantaneous component is proportional to inverse of the Drude damping rate evaluated at excitation photon energy and (ii) the relaxation rate of the exponential component is well explained by the modified two-temperature model (TTM) \cite{allen,kabanov2008} combined with ab initio calculations for the e-ph coupling strength \cite{suemoto2025}. These indicate that the luminescence dynamics in the initial and relaxation stages are governed by the e-e and e-ph scattering, respectively. It should be noted that the luminescence properties were very different for Cu and Ni, i.e., the lifetimes were short (order of 100 fs) and the intensity was two orders of magnitude lower for Ni compared to Cu. It is unclear how the luminescence properties of Cu$_{1-x}$Ni$_x$ behave for $0<x<1$. 

%The e-ph coupling of metals is related with the cooling rate of highly excited electrons \cite{allen}. Just after excitation by short light pulse, the electron has a non-thermal distribution and relaxes to thermal distribution described by Fermi-Dirac distribution via e-e scattering. Then the electron system is cooled down via e-ph scattering. 

Based on the TTM \cite{allen} and/or modified TTM \cite{kabanov2008}, the e-ph coupling constant has been estimated from the pump-probe experiments for elemental metals \cite{brorson,demsar} and high-$T_c$ cuprates \cite{kabanov2010}, whereas the effects of nonequilibrium electrons and phonons are also investigated by using more complex models \cite{mueller,waldecker,pablo2017,ono2018}. From a view point of material science, alloys are highly attractive, because we will have a chance to design the behavior of energetic electrons by controlling the composition and crystal structure. However, we can rarely find experimental or theoretical research on the ultrafast electron dynamics in alloys \cite{mcpeak2022,mcpeak2024} and it is rare to apply the TTM to the electron dynamics of alloys. 

%The low intensity of luminescence in Ni is consistent with the results of time-resolved photoemission spectroscopy (Tr-PES), because the large density-of-state (DOS) near the Fermi energy enhances the e-e scattering rate, resulting in short lifetime of energetic electrons and large non-radiative relaxation path of electrons.

In this paper, we study the composition dependence of luminescence intensity and decay rate for Cu$_{1-x}$Ni$_x$ alloys. The luminescence intensity decreases with increasing $x$ and exhibits a strong correlation with the integrated density-of-states (DOS) around the Fermi level. This reflects the extent of the phase space available for the e-e scattering. The decay rate exhibits a minimum around $x=0.2$. This unexpectedly nonmonotonic composition dependence agrees with ab initio calculations of e-ph coupling strength. Our work provides a deep understanding of physical properties of Cu-Ni alloys and opens a way to explore related phenomena in other solid solution systems. 

%%%%%%%%%%%%%%%%%%%%%%%%%%%%%%%%%%%%%%%%%%%%%%
{\it Experimental methods---}The pure metals (Cu and Ni), Cu$_{0.7}$Ni$_{0.3}$ alloy (cupronickel) and Cu$_{0.55}$Ni$_{0.45}$ alloy (constantan) were purchased from Nilaco Corporation as a form of plate. The samples with other compositions were prepared by co-melting cupronickel and Cu on a tungsten boat mounted in a vacuum chamber of a vapor deposition equipment. In order to suppress evaporation of Cu, the chamber was purged by Ar gas at a pressure of about 3 Torr. The obtained ingot was cut and polished with alumina polishing paper (\#1000---\#4000). Finally the surfaces were roughened by pressing with an alumina polishing film. The X-ray diffraction powder patters were taken for all samples including pure metals and purchased alloys. Trace of Mn was found in the purchased Cu$_{0.55}$Ni$_{0.45}$ alloy at a level of 1 \%, while no trace of metal impurity was found in cupronickel. It was verified that the lattice constant follows Vegard's law, which guarantees formation of homogeneous solid solution in whole composition range. The element analysis was made with an electron-probe micro-analyzer (EPMA) for the surfaces of metals. Traces of C, O and Al were found in melt-grown samples. Carbon is a common impurity originating from hydrocarbons in vacuum system. Oxygen and aluminum are presumed to come from the polishing alumina powder, which has practically no effect on the ultrafast luminescence properties of metals \cite{suemoto2025}.  

We used an up-conversion technique, using amplified mode-locked pulses (1.19 eV, 600 mW, 100 MHz repetition rate) from an Yb-fiber laser. The pumping pulse (typically 150 fs duration, 200 mW) was focused on the sample with an approximate spot size of 20 $\mu$m. The luminescence was collected and focused on a nonlinear optical crystal (lithium iodate, LIO) using two paraboloidal mirrors and mixed with a gating pulse. The anti-Stokes luminescence from the sample was removed by a Si-filter. The generated sum-frequency light was passed through a long-wave-pass edge filter, directed into a spectrometer consisting of tunable band-pass filters and detected by an avalanche photodiode (APD). The time resolution of this system was typically 240 fs (full width at half maximum). Detailed information of the experimental setup are presented in our previous reports \cite{suemoto2020,suemoto2025} and Supplementary Materials for Ref.~\cite{suemoto2020}. All luminescence measurements were performed in air at room temperature.

Absorptivity of the samples was measured at 1.19 eV as an increase of temperature under irradiation by the Yb-fiber laser with reduced power (20 mW). The absorptivity (emissivity) at longer wavelengths, where luminescence signal appears, was measured by using tunable light from an optical parametric amplifier (OPA) operating at a repetition frequency of 1 kHz with an average power (10 mW) to avoid damages on the sample.

As done in our previous works \cite{suemoto2019,suemoto2020,suemoto2021,suemoto2023,suemoto2025}, we first measured the absorptivity of the alloy surfaces by using calorimetry (see Fig.~S1 \cite{SM}). These spectra were used to obtain internal spectra and intensity, by normalizing the observed luminescence spectra by the energy dependence of emissivity. 

%The spectra are smooth without noticeable structures within the measured energy range. The absorptivity is lower for Cu rich alloys, because the roughness depends on the hardness of the alloy. 

%%%%%%%%%%%%%%%%%%%%%%%%%%%%%%%%%%%%
{\it Computational details---}To calculate the electron DOS, we performed electronic structure calculations for alloys. To model the crystal structure of Cu-Ni alloys, we considered a $3\times 3\times 3$ supercell of the cubic unit cell, 108 atoms in total. For each Ni concentration, we created 50 structures, where Ni atoms randomly occupy the Cu sites. We first optimized the geometry by using M3GNet, a universal interatomic potential incorporating three-body interactions \cite{m3gnet}, and determined the Ni distribution that has the lowest energy among 50 samples. 

We reoptimized the crystal structure based on density-functional theory (DFT) calculations, where the atomic positions are relaxed and the lattice constant $a$ follows the Vegard's law. In the present work, we assumed $a=(1-x) a_{\rm Cu}+xa_{\rm Ni}$, where the experimental lattice constants of $a_{\rm Cu}=3.61$ \AA \ for Cu and $a_{\rm Ni}=3.52$ \AA \ for Ni were used. The DFT calculations were performed by using the Quantum ESPRESSO (QE) \cite{qe}. The pseudopotentials of Perdew, Burke, and Ernzerhof \cite{pbe} in the pslibrary1.0.0 \cite{dalcorso} were used. The cutoff energies for the wavefunction and charge density were set to be 60 Ry and 600 Ry, respectively, and the smearing parameter was set to be 0.02 Ry \cite{smearingMV}. The $\Gamma$ point in the Brillouin zone was sampled in self-consistent field (scf) calculations, and the $k$ grid was increased to $6\times 6\times 6$ in non-scf calculations. The electron DOS was plotted with a Gaussian broadening parameter of 0.2 eV. The DOS and magnetization for various Ni content are provided in Fig.~S2 \cite{SM}. The spin splitting occurs around Ni 40\%, which is consistent with experiments (Ni 44 \%) \cite{ahern,ikeda}.

To study the e-ph coupling, we calculate the Eliashberg function given by \cite{allen}
\begin{eqnarray}
\alpha^2 F(\omega) &=& \frac{1}{\hbar N_{\rm F}N}
 \sum_{\alpha,\alpha',\bm{k},}\sum_{\beta,\bm{q}} \vert g_{\alpha,\alpha'}^{\beta}(\bm{k},\bm{q}) \vert^2
 \nonumber\\
 &\times&
 \delta (\varepsilon_{\rm F} - \varepsilon_{\alpha\bm{k}})
   \delta (\varepsilon_{\rm F} - \varepsilon_{\alpha'\bm{k}+\bm{q}})
    \delta (\omega - \omega_{\beta\bm{q}}),
    \label{eq:a2F}
\end{eqnarray}
where $\hbar$ is the Planck constant, $N_{\rm F}$ is the electron DOS at the Fermi level, and $N$ is the number of the unit cell. $\varepsilon_{\alpha\bm{k}}$ is the electron energy with the wavevector $\bm{k}$ and the band index $\alpha$, $\omega_{\beta\bm{q}}$ is the phonon frequency for the wavevector $\bm{q}$ and the branch index $\beta$, and $g_{\alpha,\alpha'}^{\beta}(\bm{k},\bm{q})$ is the matrix elements for the e-ph interaction Hamiltonian. $\omega$ is the phonon frequency. Therefore, $\alpha^2 F(\omega)$ can be regarded as the phonon DOS weighted by the e-ph matrix elements at the Fermi level. We calculate $\alpha^2 F(\omega)$ based on the density-functional perturbation theory (DFPT) \cite{dfpt} implemented in QE \cite{qe}. We considered a $2\times 2\times 2$ supercell including eight atoms in the unit cell and optimized the crystal structure. We used a $4\times 4\times 4 \ q$ grid for the phonon wave vectors, an $8\times 8\times 8 \ k$ grid for constructing the induced charge density and the dynamical matrix, and a $16\times 16\times 16 \ k$ grid for the double-$\delta$ function with the electron energy restricted at the Fermi surface. We calculated $\alpha^2 F(\omega)$ of Cu$_8$, Cu$_7$Ni$_1$, Cu$_6$Ni$_2$, and Cu$_4$Ni$_4$. The crystal structures used are provided in Supplemental Material \cite{SM}.

We calculated the e-ph coupling constant defined as
\begin{eqnarray}
\lambda\langle \omega^n \rangle = 2\int \alpha^2 F(\omega)\omega^{n-1} d\omega
\label{eq:eph}
\end{eqnarray}
with an integer $n$ and estimated the decay rate for the nonequilibrium electrons $\Gamma = 3\hbar\lambda \langle \omega^2 \rangle/(2\pi k_{\rm B}T_{\rm ph})$ with the Boltzmann constant $k_{\rm B}$ and the phonon temperature $T_{\rm ph}$ \cite{allen,kabanov2008}. 

%%%%%%%%%%%%%%%%%
\begin{figure}
\center\includegraphics[scale=0.35]{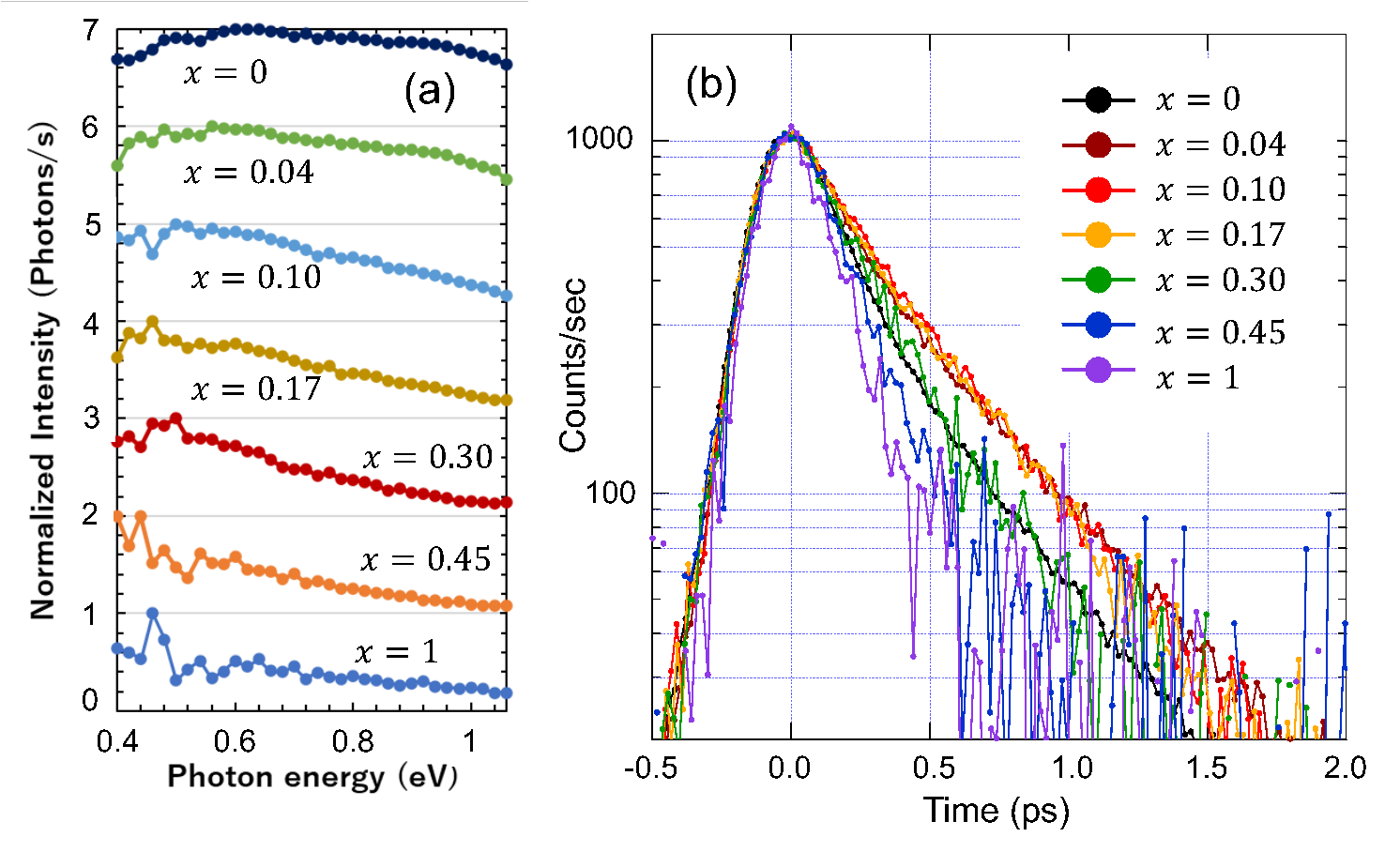}
\caption{(a) Time-resolved luminescence spectra normalized by emissivity for Cu-Ni alloys. Each spectrum is normalized at their maxima and shifted vertically. (b) Luminescence decay profiles at 0.6 eV for alloys with $x=0, 0.04, 0.10, 0.17, 0.30, 0.45$, and $1$. Each curve is normalized to 1000 at maxima. } \label{fig1} 
\end{figure}
%%%%%%%%%%%%%%%%%

%%%%%%%%%%%%%%%%%
\begin{figure*}
\center\includegraphics[scale=0.5]{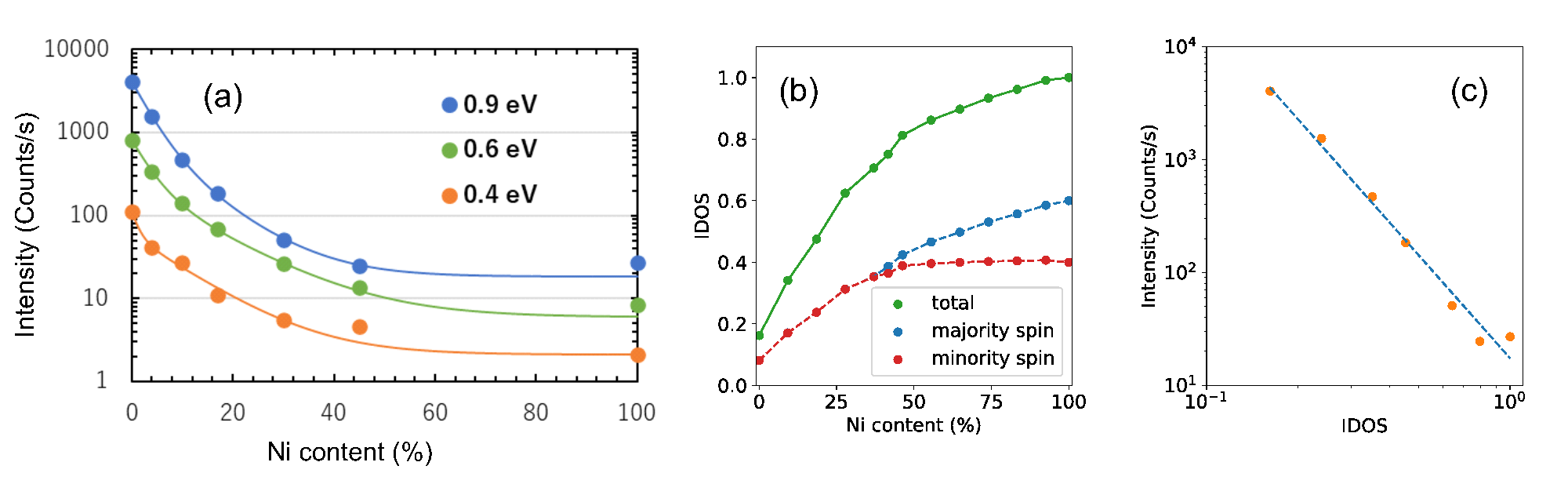}
\caption{(a) Ni concentration dependence of the internal luminescence intensity at 0.9, 0.6 and 0.4 eV. The solid curves show phenomenological fittings to the experimental data. (b) The IDOS in Eq.~(\ref{eq:IDOS}) as a function of the Ni content. The spin splitting occurs around 40 Ni\%. $\varepsilon_{\rm min}$ and $\varepsilon_{\rm max}$ were set to be $-1.2$ and 0.4 eV, respectively. The IDOS is normalized by its maximum. (c) Dependence of the internal luminescence intensity on IDOS. The dashed curve is an exponential fit with $17.4\times$IDOS$^{-3.03}$. }\label{fig2}
\end{figure*}
%%%%%%%%%%%%%%%%%

%%%%%%%%%%%%%%%%%
\begin{figure}
\center\includegraphics[scale=0.45]{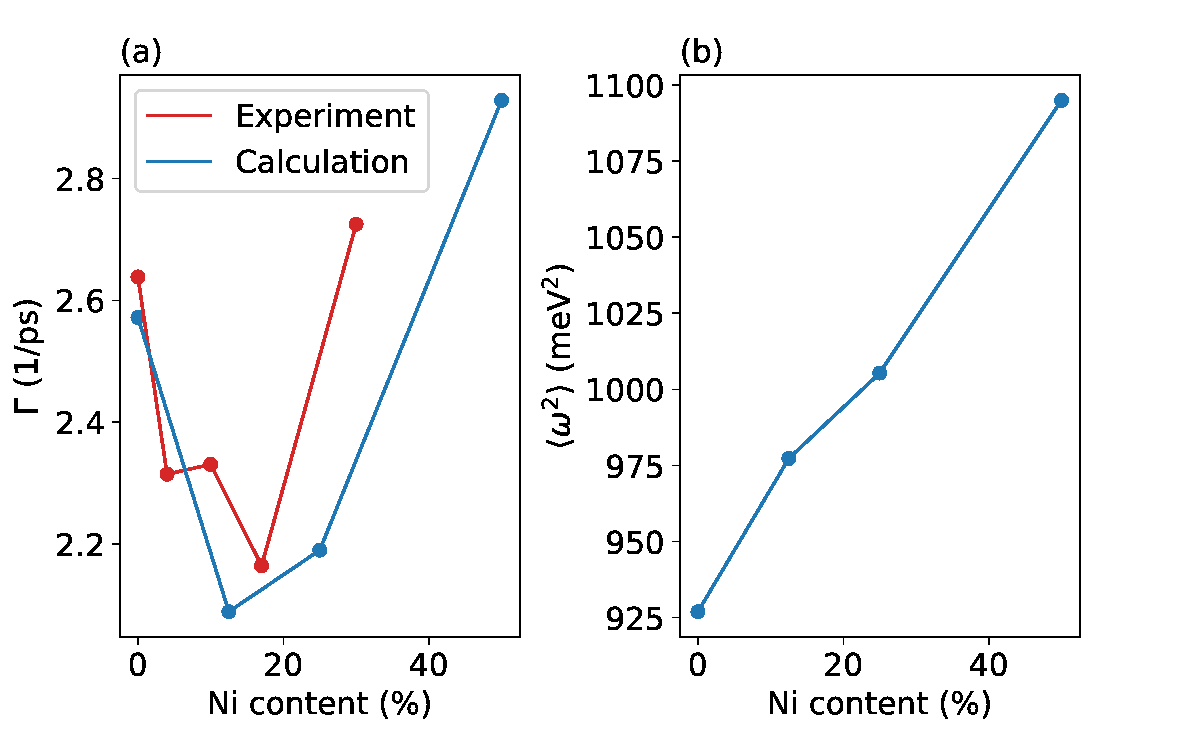}
\caption{(a) Luminescence decay rate at 0.6 eV as a function of Ni content. Red and blue circles indicate the experimental and calculated data, respectively. The factor 1.7 is multiplied to the calculated $\Gamma$. (b) The second moment of the phonon spectra as a function of Ni content. } \label{fig3} 
\end{figure}
%%%%%%%%%%%%%%%%%

%%%%%%%%%%%%%%%%%%%%%%%%%%%%%%%%%%%%%%%%%%%%%%
{\it Luminescence intensity---}Figure \ref{fig1}(a) shows the internal time-resolved luminescence spectra measured at time origin. The abscissa is proportional to photon numbers per second per unit energy interval and the spectra are normalized at maximum of each curve. The spectra for pure Cu and Ni are basically in agreement with the sand-blasted samples in our previous publication \cite{suemoto2025}. The slight deviations are ascribed to the difference in the roughness and the excitation power density, which depends on the optics of up-conversion measurement equipment. We can clearly see that the weight of the spectrum moves to lower energy side by increasing Ni content from $x=0$ to $x=0.45$. The spectrum for Ni 45\% alloy is almost equal to that of pure Ni. This result indicates that the electron temperature, or average excess kinetic energy of electrons is lower in Ni rich alloys. 
 
The composition dependences of the internal luminescence intensity evaluated at $t=0$ ps are plotted for 0.9, 0.6 and 0.4 eV in Fig.~\ref{fig2}(a). For 0.9 eV, the intensity decreases to 1/160 when the Ni content is increased from 0 to 45\%, while the decrease is milder for 0.6 and 0.4 eV. 

In the previous work, we have demonstrated that the luminescence intensity is correlated with the inverse of the Drude damping rate \cite{suemoto2025}. This is based on a phenomenological model for the complex interplay between the surface plasmon polariton and hot electrons. However, the estimation of the Drude damping rate for Cu-Ni alloys is difficult due to the lack of the frequency-dependent optical constants. We believe that the underlying physics behind the instantaneous response is governed by the e-e scattering within the energy window around the Fermi level, where electrons with different energies are scattered to unoccupied states at elevated electron temperature conditions. Alternatively, we calculate the integrated DOS 
\begin{eqnarray}
 {\rm IDOS}(\varepsilon_{\rm min},\varepsilon_{\rm max}) = \int_{\varepsilon_{\rm min}}^{\varepsilon_{\rm max}} \left[ D_\uparrow(\varepsilon) + D_\downarrow(\varepsilon)\right] d\varepsilon,
 \label{eq:IDOS}
\end{eqnarray}
where $D_\sigma(\varepsilon)$ is the electron DOS with energy $\varepsilon$ and spin $\sigma$. This is a rough estimate of the number of the single-particle states involving the e-e scattering events in our excitation condition. We assume $\varepsilon_{\rm min} = -1.2$ eV that is nearly equal to the photon energy of pump pulse (1.19 eV) with opposite sign and treat $\varepsilon_{\rm max}$ as a parameter accounting for the number of excited electrons. We have considered $\varepsilon_{\rm max}=0.0, 0.4, 0.8$, and $1.2$ eV and show our analysis for $\varepsilon_{\rm max}=0.4$ eV below.

As shown in Fig.~\ref{fig2}(b), IDOS increases with the Ni content and the increase becomes moderate at higher Ni content. Figure \ref{fig2}(c) shows the observed luminescence intensity as a function of the calculated IDOS. We can see a good correlation between them, i.e., the intensity shows an IDOS$^{-3.03}$ dependence. Note that the integrand of the e-e collision term in the Boltzmann transport equation is proportional to the cube of the electron DOS \cite{ono2020}, and the e-e collision accelerates the electron thermalization \cite{mueller,ono2018}. In addition, the luminescence intensity reflects the initial relaxation before the energy transfer to phonons \cite{suemoto2025}. Therefore, our model calculations demonstrate that the integrated value of the collision term of the Boltzmann equation is well approximated by the power function of IDOS, and allow to estimate the relevant energy window governing the relaxation process. When $\varepsilon_{\rm max}$ is changed from 0 to 1.2 eV, the power law changes from $-2.92$ to $-3.49$ that are equal to $-3$ within an error of 17 \% (see Supplemental Material \cite{SM}). More quantitative discussion requires solving the Boltzmann equation, but such a study is beyond the scope of the present work.

%Therefore, the higher the DOS, the faster the e-e thermalization (i.e., the weaker the luminescence intensity).

{\it Luminescence decay---}Time development of luminescence intensity was investigated at 0.6 eV and shown in Fig. \ref{fig1}(b), where the curves are normalized to 1000 at their maxima. In contrast to the monotonic concentration dependence of luminescence intensity at 0.9 eV, it shows a complicated behavior. The slope between 0 and 0.5 ps decreases from pure Cu by adding Ni and keeps similar value from Ni 4\% to 17 \% (i.e., the relaxation is slowed). Above 17 \%, it starts to increase and reaches maximum value at Ni 100\% (i.e., the relaxation becomes faster). To analyze this behavior quantitatively, we decomposed the decay profile into an instantaneous response and an exponentially decaying component \cite{suemoto2025}. The decay profiles are well fitted by a sum of a delta function and an exponential function with convolution by the instrumental response function. 

The decay rates $\Gamma_{\rm exp}$ of the exponential component at 0.6 eV are shown in Fig.~\ref{fig3}(a). By adding Ni to pure Cu, $\Gamma_{\rm exp}$ decreases first till Ni 17\% and turns to increasing over 17\%. We also plotted the calculated $\Gamma$.  To correct the underestimation for the decay rate, the factor 1.7 was multiplied to $\Gamma$ \cite{suemoto2025}. It is clear that the calculated $\Gamma$ is nonmonotinic function for Ni content. This shows that the luminescence decay rate corresponding to the cooling rate is dominated solely by e-ph coupling and that the disorder originating from alloying has minor contribution to the decay process.  

To understand the origin of the nonmonotonic behavior, we calculated the second moment of the phonon spectra, i.e., $\langle \omega^2 \rangle = 2\int F(\omega) \omega d\omega$ for Cu-Ni alloys, where $F(\omega)$ is the phonon DOS. The $\langle \omega^2 \rangle$ monotonically increases with Ni content, as shown in Fig.~\ref{fig3}(b). One should notice that the $\langle \omega^2 \rangle$ is obtained if $\alpha^2 F(\omega)$ is replaced with $F(\omega)$ in Eq.~(\ref{eq:eph}). The nonmonotonic behavior is not due to the redistribution of the phonon spectra, but due to the weakened $\alpha^2 F(\omega)$. We speculate that in dilute alloys the matrix elements $\vert g\vert^2$ and/or the phase space for the e-ph scattering are decreasing function of Ni concentration, resulting in an anomalous decrease in $\lambda \langle \omega^2 \rangle$. The $\alpha^2 F(\omega)$ and the phonon DOS for Cu-Ni alloys are provided in Supplemental Material \cite{SM}. 

% (e.g., the electron screening becomes strong \cite{smith})

%the Cu$_7$Ni$_1$, Cu$_8$, and Cu$_7$Zn$_1$ in Fig.~\ref{fig3}(b). When one of Cu is replaced with Ni, the highest frequency increases from 30 meV (Cu$_8$) to 31 meV (Cu$_7$Ni$_1$) because Ni is lighter than Cu in weight. For pure Cu, strong peaks at lower and higher frequencies correspond to the zone-boundary phonons in the transverse and longitudinal acoustic branches, respectively. However, these peaks are weakened for small Ni concentrations. The calculated $\alpha^2F(\omega)$ for other Cu-Ni models are provided in the Supplemental Material \cite{SM}. 

We calculated $\lambda\langle \omega^2 \rangle$ for Cu$_7Z_1$ with period 4 metals $Z$ from K to Zn. Among them, Cu$_7$Ni$_1$ has the smallest value of $\lambda\langle \omega^2 \rangle$. We also calculated $\lambda\langle \omega^2 \rangle$ for $X_7Y_1$ with $X=$ Cu, Ag, Au and $Y=$ Ni, Pd, Pt, Cu, Ag, Au. A small inclusion of group 10 atoms always gives rise to a decrease in $\lambda\langle \omega^2 \rangle$. The smallest $\lambda\langle \omega^2 \rangle$ is 16.1 meV$^2$ for Au$_7$Pt$_1$. Our systematic investigation suggests that the relaxation dynamics can be manipulated by alloying. These results are provided in Supplemental Material \cite{SM}. 

%It should be noted that $\lambda\langle \omega^2 \rangle$ is increased in Cu$_7$Zn$_1$ (see Fig.~\ref{fig3}(b)). This is due to the enhanced $\alpha^2F(\omega)$ near the one-boundary phonon energies. 

%The nonmonotonic relationship between the e-ph coupling constant and the composition ratio in alloys is a universal behavior in noble metal-based solid solutions. 

%%%%%%%%%%%%%%%%%
%\begin{table}\begin{center}\caption{Lattice constants of RbNi$_3$S$_4$. The models 1 and 2 are calculated within GGA+vdW. }
%{\begin{tabular}{lrrr}\hline\hline
%--------------------------------------------------------------------------------
%\hspace{3mm} & $a$ (\AA) \hspace{3mm} &  $b$ (\AA) \hspace{3mm} & $c$ (\AA) \\
%\hline
%Exp. (present work) \hspace{3mm} & 5.57 \hspace{3mm} & 9.66 \hspace{3mm} & 13.16   \\
%Model 1  \hspace{3mm} & 5.701 \hspace{3mm} & 9.679 \hspace{3mm} & 14.146  \\
%Model 2  \hspace{3mm} & 5.613 \hspace{3mm} & 9.647 \hspace{3mm} & 13.146  \\
 %--------------------------------------------------------------------------------
%\hline
%\end{tabular}}\label{table2}\end{center} \end{table}
%%%%%%%%%%%%%%%%%

%%%%%%%%%%%%%%%%%%%%%%%
{\it Conclusion---}Luminescence intensity and decay profiles were measured in Cu-Ni alloys and the composition dependence was compared with a theoretical prediction. Nonmonotonic composition dependence of the decay rate was found below $x=0.2$. Elongation of the luminescence lifetime by adding transition metal to noble metal is a rather unexpected result, because the increase of DOS near the Fermi level accelerates relaxation due to e-ph scattering. We emphasize that the quantitative prediction of the electron relaxation dynamics will promote alloy design for desired electronic and optical properties, e.g. to realize longer lifetime.
Size effect in nanocrystals may help to modify the Eliashberg function to further reduce the e-ph interaction strength, leading to improved quantum efficiency of nanoparticle imager. The inverse effect, i.e., an increase in the e-ph coupling by foreign atom doping, may enhance the superconducting transition temperature of metals. 

We also found that the smaller the IDOS around the Fermi level, the larger the luminescence intensity. This may reflect the phase space available for the e-e scattering. This concept will be useful for controlling the brightness of alloys.

\begin{acknowledgments}
This work was supported by JSPS KAKENHI (Grant Number JP20K03823). T.S. is indebted to the Institute for Solid State Physics at the University of Tokyo for its support with the Joint Research Program. Some of the numerical calculations were done using the facilities of the Supercomputer Center, the Institute for Solid State Physics, the University of Tokyo, and the supercomputer “MASAMUNE-IMR” at the Center for Computational Materials Science, Institute for Materials Research, Tohoku University. 
\end{acknowledgments}

%%%%%%%%%%%%%%%%%
%\appendix
%%%%%%%%%%%%%%%%%%%%%%%

%===================================================================%
%   References
%===================================================================%

%\bibliography{refs}

\end{document}